# Solar cells and thermoelectric generators as finite-time chemical heat engines

Tom Markvart*
School of Engineering Sciences, University of Southampton, UK, and
Centre for Advanced Photovoltaics, Czech Technical University in Prague

**ABSTRACT** This paper shows a fundamental thermodynamic similarity between thermoelectric and photovoltaic energy converters which, at open circuit, can be represented as isochoric engines generating a finite chemical potential which appears as voltage at the terminals of the device. We show that, allowing for the temperature variation of the Seebeck coefficient, the maximum energy efficiency is intrinsically lower than the Carnot efficiency, as assumed in most of the literature, although more sophisticated strategies may exist to recuperate this loss. At finite current, further losses can be modelled in terms of finite-time thermodynamics.

Thermodynamic theories of thermoelectric generators continue to attract the interest of the research community (see, for example, [1-6]). Observed for the first time more than two hundred years ago [7,8], the thermoelectric phenomena were explained in terms of classical thermodynamics [9], later refined with the understanding of irreversible thermodynamics [10, 11]. The first observation of the photovoltaic effect dates to a similar time as the discovery of thermoelectricity [12] but practical applications of solar cells and thermoelectric generators had to await the advent of semiconductors in the mid 20th century.

In this note we demonstrate a fundamental similarity between the thermoelectric and photovoltaic generators pointing out, at the same time, inherent irreversibilities and losses in the conversion process. We show that, at open circuit, both types of converters work as isochoric chemical engines, where electrons or photons as the working medium generate a finite chemical potential equal to the output voltage. Extraction of current from the generator and the resulting losses are naturally modelled in the language of finite-time thermodynamics.

To this end, we consider the thermoelectric generator at open circuit, as an energy (rather than power) converter, transforming heat into the electrostatic energy $qV_{oc}$ of charge carriers at the contacts. The generator legs are formed by n and p type semiconductors, with sufficient density of doping and high enough bandgap to limit the conductivity to a single type of charge carrier in each leg (Fig.1). The conversion process is described by Kelvin's relations [13]

$$\frac{d\pi}{dT} = \varepsilon + \tau \qquad (1)$$

$$\pi = \varepsilon T \qquad (2)$$

where $\pi = \pi_p - \pi_n$, $\varepsilon = \varepsilon_p - \varepsilon_n$ and $\tau = \tau_p - \tau_n$ are the Peltier, Seebeck and Thomson coefficients. The Seebeck and Thomson coefficients are related by

$$\tau = T\frac{d\varepsilon}{dT} \qquad (3)$$

The difference $\pi(T_h) - \pi(T_c)$ of heat flowing in and out of the converter can be written as

$$\pi(T_h) - \pi(T_c) = \int_{T_c}^{T_h}\frac{d\pi}{dT}dT = \int_{T_c}^{T_h}\varepsilon(T)dT + \int_{T_c}^{T_h}T\frac{d\varepsilon}{dT}dT \qquad (4)$$

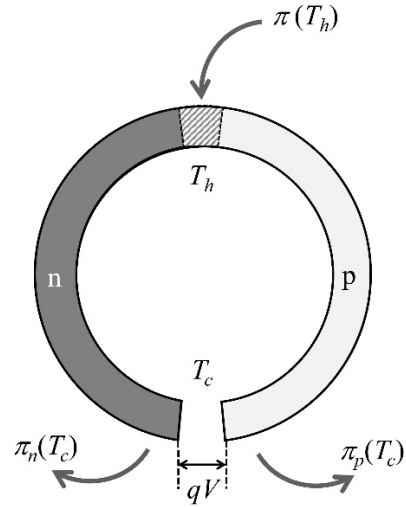

Fig. 1. A schematic diagram of the thermoelectric generator. Heat is absorbed at a junction at temperature $T_h$ between p and n type semiconductors, and rejected at contact at temperature $T_c$ ($<T_h$), representing a cold junction.

*tm3@soton.ac.uk

The integral over $\varepsilon$ is equal to the voltage generated by the thermoelectric generator:

$$V_{oc} = \int_{T_c}^{T_h} \varepsilon(T)dT \qquad (5)$$

A thermodynamic interpretation is provided by linking the Seebeck coefficients to the carrier entropies $s_n$ and $s_p$ [2, 14]:

$$\varepsilon = \frac{1}{q}(s_p - s_n) = \frac{1}{q}s \qquad (6)$$

where $q$ is the elementary charge. Equation (4) can then be written in the following alternative form:

$$qV_{oc} = \{\pi(T_h) - \pi(T_c)\} - \int_{T_c}^{T_h} T\frac{ds}{dT}dT \qquad (7)$$

which, with the help of (2), can be transformed into

$$qV_{oc} = \left(1 - \frac{T_c}{T_h}\right)\pi(T_c) - T_c\sigma_i \qquad (8)$$

where

$$\sigma_i = \frac{1}{T_c}\int_{T_c}^{T_h} Tds - \{s(T_h) - s(T_c)\} = \frac{1}{T_c}\int_{T_c}^{T_h}(T - T_c)ds \qquad (9)$$

Equation (8) has a clear interpretation. Noting that $\sigma_i \geq 0$, the heat $\pi(T_h)$ flowing into the high-temperature junction is converted into voltage with the Carnot efficiency $(1-T_c/T_h)$, less "lost work" equal to $T_c\sigma_i$, where $\sigma_i$ is the entropy generation in the conversion process. In mechanical systems, Eq. (8) is sometimes called the Gouy-Stodola relation [15].

In much of the literature on thermoelectric generation, the irreversible entropy production (9) is absent. This is usually due to neglecting the Thomson heat or assuming a constant Seebeck coefficient which results in constant entropy $s$ during the conversion process and, consequently, zero entropy generation. It was suggested [16] that this adiabatic cooling can be more rigorously achieved by nonuniform doping which would keep the thermopower constant during cooling.

Equation (9) can be transformed further within "per carrier" thermodynamics where the first law for carriers in either of the two legs becomes (see SI)

$$du = Tds - pdv \qquad (10)$$

where $v = 1/n$ and $u$ are the volume and energy per carrier, and $n$ is the (majority) carrier density, assumed equal to the doping density. Focusing on usual thermoelectric generators where the doping in the two legs of the thermocouple is uniform and the volume $v$ is constant, we obtain $du = Tds$. Equation (9) then becomes

$$\sigma_i = \frac{1}{T_c}\Delta u - \Delta s \qquad (11)$$

*tm3@soton.ac.uk

where $\Delta u$ and $\Delta s$ are the energy and entropy differences for carriers at the hot and cold junctions.

Equation (11) gives a characteristic entropy generation during an isochoric process observed, for example, in the Stirling engine. The first term on the right-hand side is equal to an increase of entropy of the low-temperature reservoir by absorbing energy $\Delta u$ from the thermoelectric, and the second term is the charge-carrier entropy change in the converter.

We note in passing that Eq. (11) can be written as,

$$T_c\sigma_i = \int_{T_c}^{T_h}\left(1 - \frac{T_c}{T}\right)du \qquad (12)$$

Equation (12) pictures the lost work due to irreversibility in a particularly clear form, by equating it to the reversible work that could be (but isn't) carried out by converting infinitesimal energy amounts $dU$ when connected to energy reservoirs at an appropriate temperature, equal to the temperature of the working medium.

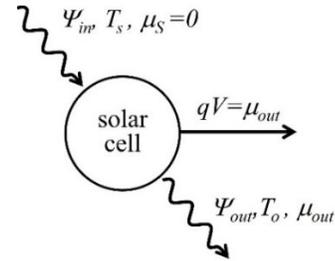

Fig. 2. A schematic diagram of the solar cell, represented as a converter of light beams described by a photon flux, temperature and chemical potential, generating voltage $V$.

This pattern of voltage production can be compared with photovoltaics (Fig. 2). The conversion process in solar cells can be represented as a transformation of an incident beam of light into a beam emitted by the solar cell where photons have a finite chemical potential $\mu_{out}$ equal to the voltage at the terminals of the solar cell [17-19]. In this process, the size of the light beam (the étendue) plays the role of volume, and the photon flux the role of particle (carrier) density in the thermodynamic description. At open circuit, the conversion under the maximum concentration of sunlight when the incident and emitted beam are of the same size then corresponds to an isochoric process at constant volume per carrier considered earlier for the thermoelectric generator.

We now use the Maxwell relation (see, for example, [14, 20])

$$\left(\frac{\partial \mu}{\partial T}\right)_{V,N} = -\left(\frac{\partial S}{\partial N}\right)_{V,T} = -s \quad (13)$$

where $s$, in this instance, is the entropy per photon. Integrating and perceiving the incident solar beam as blackbody radiation at temperature $T_S$ and chemical potential zero gives

$$\mu_{out} = \int_{T_o}^{T_S} s\, dT \quad (14)$$

where $T_o$ is the temperature of the solar cell. Expression (14) for photons is analogous to (5) for charge carriers in a thermoelectric generator.

The parallel with thermoelectric conversion can be pursued further by writing

$$qV_{oc} = \mu_{out} = u_{out} - T_o s_{out} \quad (15)$$

where the photon energy $u_{out}$ and entropy $s_{out}$ refer to the emitted beam. A simple manipulation using

$$\mu_S = u_S - T_S s_S = 0 \quad (16)$$

for the incident solar beam gives

$$qV_{oc} = \left(1 - \frac{T_o}{T_S}\right)u_S - T_o \sigma_i \quad (17)$$

where

$$\sigma_i = \frac{u_S - u_{out}}{T_o} - (s_S - s_{out}) \quad (18)$$

can be seen as the analogues of (8) and (11) for thermoelectric generators.

The voltage loss $T_o \sigma_i$ through "thermalization" represents the principal efficiency loss in solar cells. This is in contrast with thermoelectric generators where this loss is small and frequently neglected. The reason for this different magnitude in the two applications lies in the temperature difference $T_h - T_c$: several thousand degrees for solar cells but only several hundred for thermoelectric generators. Since $\sigma_i$ is proportional to $(T_h - T_c)^2$, one would expect two orders of magnitude smaller loss in a thermoelectric in comparison with a solar cell.

Considerable research activity exists to reduce or eliminate this loss in "hot carrier solar cells" [23 -26]. The thermodynamic view suggests that, by decoupling carrier thermalization from photons, cooling of carriers can be carried out adiabatically, in a similar strategy to thermoelectric generators [16].

Solar cells and thermoelectric generators are not, of course, energy converters but power converters. The effect of current generation on voltage can be displayed by way of voltage-current characteristic, in a form common in electrochemistry. The quasi-static energy / voltage generation at open circuit can be parallel the operation of a heat engine in classical thermodynamics. With the passage of fine current, voltage will be reduced by entropy generation, as predicted by finite-time (irreversible) thermodynamics [5, 21, 22].

The generated voltage can then be written as

$$qV = qV_{oc} - (T_c \sigma_{kin} + IR) \quad (19)$$

and power determined as the product $VI$. The term $T_o \sigma_{kin}$ is the entropy generation associated with the finite rate of turnover of the isochoric engine. In solar cells, $\sigma_{kin}$ is the difference in photon entropy in the incident (photon flux $\Psi_{in}$) and emitted (photon flux $\Psi_{out} = \Psi_{in} - I/q$ ) light beams, equal to

$$\sigma_{kin} = k_B \ln\left(\frac{\Psi_{in}}{\Psi_{out}}\right) = k_B \ln\left(\frac{q\Psi_{in}}{q\Psi_{in} - I}\right) \quad (20)$$

In thermoelectrics, this term would arise from a finite rate (and therefore irreversible) of heat exchange between the generator and the two thermal reservoirs. If the loss (9) is neglected, the thermoelectric generator becomes an endoreversible heat engine [6].

The second term in brackets on the right hand side of Eq. (19) is due to resistive losses in the two legs of the thermoelectric generator, or in the quasi-neutral charge collection regions in solar cells. In solar cells, the resistance $R$ appears as the series resistance in the current-voltage characteristic which, in good solar cells, is small.

In conclusion, we have shown that there is a fundamental similarity between energy/voltage production in thermoelectric generators and solar cells which, in standard devices, can be pictured as isochoric chemical heat engines. The isochoric nature implies a thermalization loss by irreversibility which is usually small in thermoelectrics but the main efficiency loss in solar cells. Extraction of current moves the conversion process into the realm of finite-time irreversible thermodynamics.

**Acknowledgement**
This work was supported by project "Energy Conversion and Storage", grant no. CZ.02.01.01/00/22_008/0004617, programme Johannes Amos Commenius, Excellent Research.

*tm3@soton.ac.uk

*tm3@soton.ac.uk

SUPPLEMENTARY INFORMATION

**Derivation of Eq. (10)**

We start with an expression for the chemical potential in the form

$$\mu = \left(\frac{\partial F}{\partial N}\right)_{V,T} = u - Ts \qquad (S1)$$

where $F$ is the Helmholtz free energy, $u = (\partial U / \partial N)_{V,T}$ and $s = (\partial S / \partial N)_{V,T}$ are the energy and entropy per particle and other symbols have their usual meaning. The Gibbs-Duhem equation

$$d\mu = -s_p dT + vdp \qquad (S2)$$

where $s_p = S/N$ is subjected to Legendre transformation

$$d\mu = -s_p dT + vdp + d(vp) - d(vp) = \left\{\frac{d(vp)}{dT} - s_p\right\} dT - pdv \qquad (S3)$$

and the term in braces, equal to $(\partial \mu / \partial T)_v$ is identified as the negative of the entropy per carrier in Eq. (S1) using Eq. (13) of text. Taking the differential of (S1) and combining with (S3) gives Eq. (10) in text.